\documentclass[aps,prb,preprint,showpacs,preprintnumbers,amsmath,amssymb]{revtex4}
\usepackage{graphicx}
\usepackage{float}
\begin{document}
\title{ Engineering semiconductor hybrids for sensing  }

\author{Godfrey Gumbs$^{1,2}$, Andrii Iurov$^{1}$ and Danhong Huang$^3$}
\affiliation{$^{1}$Department of Physics and Astronomy, Hunter College of the
City University of New York, 695 Park Avenue, New York, NY 10065, USA\\
$^{2}$ Donostia International Physics Center (DIPC),
P de Manuel Lardizabal, 4, 20018 San Sebastian, Basque Country, Spain\\
$^{3}$Air Force Research Laboratory, Space Vehicles Directorate, Kirtland Air Force Base, NM 87117, USA}

\date{\today}

\begin{abstract}

The effect of screening of the coulomb interaction between two layers of
two-dimensional electrons, such as in graphene, by a highly doped
semiconducting substrate is investigated. We employ the random-phase approximation to
calculate the dispersion equation of this hybrid structure in order to determine the
plasmon excitation spectrum.
When an electric current is passed through   a layer,   the
low-frequency plasmons in the layer may bifurcate into separate streams due to
the current-driving effect. At a critical wave vector, determined by the separation between
layers   and their distance from the surface, their phase velocities
may be in opposite  directions and a surface plasmon instability leads to the emission
of radiation.  Applications to detectors and
other electromagnetic devices exploiting nano-plasmonics are discussed.
\end{abstract}
\vskip 0.2in
\maketitle

\section{Introduction and formalism}

Rapid progress  has been made in recent years  in combining two
or more different types of materials to form hybrid nanoparticles
on the same nanosystem\,\cite{new1,new2,new3,new4,new5,new6,new7,new8,new9,new10}.
The main goal of these
endeavors  is to produce unusual materials with desired functionality
such as  sensors. Examples include  inorganic cage structures
obtained by a selective edge growth mechanism of the metal onto the
semiconductor nanocrystal.
Successful approaches now exist for synthesizing composites with a high
degree of control over their shapes, compositions and interfacial properties.
The  advances in architecture are very impressive and  the resulting
advanced materials are now being  investigated
for application of hybrid nanoparticles in electronic devices such as a
microlaser,  optical components, catalysis,  and photocatalysis.
A necessary condition for  efficient  operation of these
composite structures is electronic coupling across the constituent
interfaces.
\medskip

\begin{figure}[t]
\centering
\includegraphics[width=0.45\textwidth]{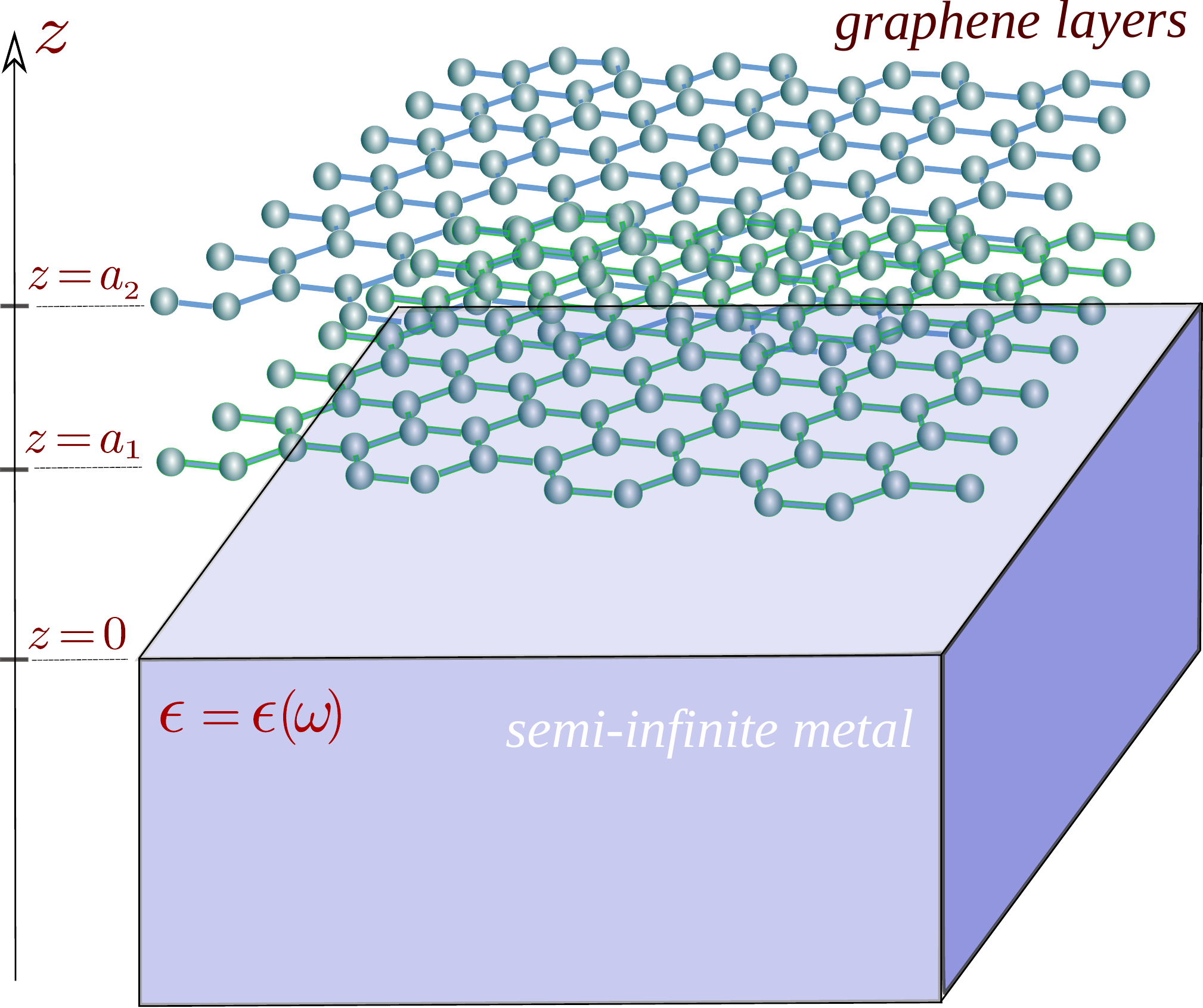}
\caption{(Color online)  Schematic illustration of a spiler generator consisting of a thick (semi-
infinite) conducting material on whose surface a plasmon resonance may induce  an instability by coupling
to the 2D plasmons on a pair of thin layers such as graphene, silicene or 2DEG at a hetero-interface.}
\label{FIG:1}
\end{figure}

This paper focuses on a doped two-dimensional (2D)
semiconductor layer hybrid nanoparticle structure  for which there is
a high degree of control over its performance as a source of electromagnetc
radiation.  We point out some challenges for further  understanding and development
of semiconductor-metal hybrid nanoparticles.
Specifically, we  consider a composite nano-system as shown in Fig.[\,\ref{FIG:1}]
consisting of a thick layer of conducting substrate on whose surface the plasmon
is activated in proximity with a pair of thin layers. We demonstrate how the
screening of the Coulomb coupling of the plasmons in this pair of layers
by the charge density fluctuations on the surface of a semi-infinite substrate
affects the surface plasmon instability that leads to the emission of radiation (spiler).
The excitation of these plasmon modes is induced by resonant external optical fields.
As an emitter, the spiler may be activated  optically.
Spiler exploits the
flexibility of choosing its constituents to produce coherent sources
of radiation. Applications to photodetectors and electromagnetic-wave devices
exploiting nano-plasmonics are explored.
\medskip

Our approach models an ensemble consisting of a pair of 2D layers
and a thick layer of a conducting medium that emits radiation when an
electric field  splits the plasmon spectrum  which results in an
instability  as the phase velocities associated with these
plasmon branches have opposing signs at a common frequency.
\medskip

In our formalism, we consider a nano-scale system consisting
of a pair of 2D layer structure with a double layer positioned
at $z=a_1$ and $z=a_2$  ($0<a_1<a_2$)
and a thick conducting material.  The
layer may be monolayer graphene or a 2D electron gas (2DEG), such as a
semiconductor  inversion layer or a high electron mobility
transistor. The graphene layer may have a gap, thereby
extending the flexibility of the composite system that
also incorporates a thick layer of conducting material as
depicted in Fig.\,\ref{FIG:1}. The excitation spectra
of allowable modes will be determined from the solutions of the dispersion equation
which includes the screened Coulomb coupling between layers and between each layer and
the semi-infinite substrate. Following standard procedures within the random-phase approximation (RPA)\,\cite{Arxiv, Tso},
calculation shows that the plasmon dispersion
equation is $S_{C}^{(2)}(q_{||},\omega)=0 $, where

\begin{eqnarray}
&&S_c^{(2)}(q_{||},\omega) =\left\{  1+  \frac{2\pi e^2}{\epsilon_s q_{||}}\,\Pi_{2D;2}^{(0)} (q_{||},\omega)
\left [  1  +  e^{-2q_{||} a_2 }\,\frac{\omega_p^2}{2\omega^2-\omega_p^2} \right]
\right\}
\nonumber\\
&\times&
\left\{  1+  \frac{2\pi e^2}{\epsilon_s q_{||}}\,\Pi_{2D;1}^{(0)} (q_{||},\omega)
\left [  1  +  e^{-2 q_{||} a_1 }\,\frac{\omega_p^2}{2\omega^2-\omega_p^2} \right]
\right\}
\nonumber\\
&-&   \left(  \frac{2\pi e^2}{\epsilon_s q_{||}} \right)^2
\Pi_{2D;1}^{(0)} (q_{||},\omega)\,\Pi_{2D;2}^{(0)} (q_{||},\omega)
\left [  e^{-q_{||} |a_1-a_2|}   +  e^{- q_{||} (a_1 +a_2) }\,\frac{\omega_p^2}{2\omega^2-\omega_p^2}
\right]^2\ ,
\end{eqnarray}
whose $\Pi_{2D}^{(0)} (q_{||},\omega)$ is the 2D polarization function  and
$\omega_p$ is the bulk plasma frequency of the semi-infinite medium.
In our numerical calculations, we employ  $\Pi_{2D}^{(0)} (q_{||},\omega)\approx Cq_\parallel^2/\omega^2$.
For a 2DEG, we have $C=n_{2D}/m_{2D}^\ast$ with electron density $n_{2D}$ and
electron effective mass $m^\ast$.
For graphene, we find

\begin{equation}
C= \frac{2\mu}{\pi\hbar^2}\left\{ 1-\frac{\Delta^2}{\mu^2} \right\}\ ,
\end{equation}
where $\mu$ is the chemical potential and $\Delta$ is the induced gap
between valence and conduction bands.
Then,  we obtain a quartic  equation in $\omega^2$. The plasmon frequencies in this approximation are

\begin{figure}
\centering
\includegraphics[width=0.5\textwidth]{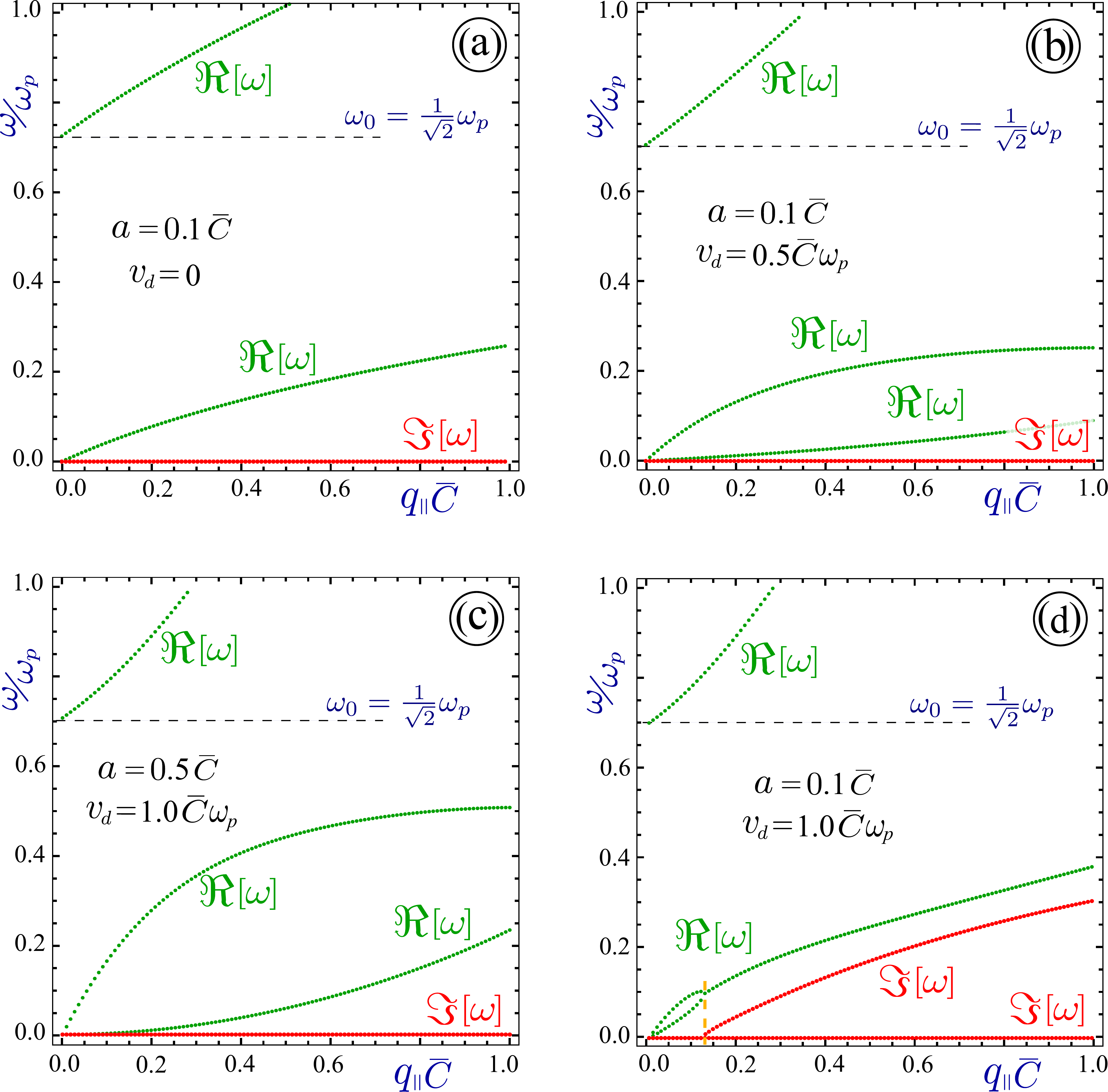}
\caption{(Color online) Complex frequencies yielding the plasmon dispersion
(real part) and inverse lifetime (imaginary part) for a 2D layer
and a semi-infinite dielectric medium as presented in Fig.\,\ref{FIG:1}. Panel $(a)$
demonstrates the case when there is no drift current, whereas panels $(b)$, $(c)$ and $(d)$
display the situations with a finite drift velocities $v_d=0.5\,\bar{C}\omega_p$,
$1.0\,\bar{C}\omega_p$ and $1.0\,\bar{C}\omega_p$. Panel $(c)$ corresponds to $a=0.5\,\bar{C}$ for the
distance between the layer and the surface, while
the three other panels $(a)$, $(b)$ and $(d)$  to $a=0.1\,\bar{C}$.}
\label{FIG:a1}
\end{figure}

\begin{eqnarray}
&& \Omega_{1}(q_\parallel)/\omega_p = 1/\sqrt{2} + q_\parallel\,(\overline{C}_1+\overline{C}_2)/\sqrt{2} + \mathcal{O}[q_\parallel^2]\ , \\
\nonumber
&& \Omega_{2}(q_\parallel)/\omega_p =  q_\parallel \sqrt{\overline{C}_1 a_1 + \overline{C}_2 a_2 + \sqrt{\mathcal{A} }}+ \mathcal{O}[q_\parallel^2]\ , \\
\nonumber
&& \Omega_{3}(q_\parallel)/\omega_p =  q_\parallel \sqrt{\overline{C}_1 a_1 + \overline{C}_2 a_2 - \sqrt{\mathcal{A} }} + \mathcal{O}[q_\parallel^2]\ ,
\label{solutions}
\end{eqnarray}
where $\mathcal{A} \equiv (\overline{C}_1 a_1 -\overline{C}_2 a_2)^2 + 4 \overline{C}_1 \overline{C}_2 a_1^2$.
Here, in  our notation, $\overline{C}_j=2\pi e^2C_j/(\epsilon_s\omega_p^2)$ for $j=1,2$.
The spectral function yields  real frequencies. A 2D layer  interacting with
the half-space has two resonant modes. Each pair of 2D layers interacting in isolation
far from the half-space conducting medium supports a symmetric and an
antisymmetric mode\,\cite{DasSarma}.  In the absence of a driving current, the analytic solutions  for the plasmon modes for a
pair of 2D layers that are Coulomb coupled to a half-space are given by
the term $4 \overline{C}_1 \overline{C}_2 a_1^2$ which plays the role of ``Rabi coupling''.
Clearly, for long wavelengths, only $\Omega_1(q_\parallel)$ depends on $\omega_p$.\
However, the excitation spectrum changes dramatically when a current is driven
through the configuration.  Under a constant electric field, the carrier distribution
is modified, as may be obtained by employing the relaxation time approximation
in the equation of motion for the center-of-mass momentum. For carriers in a parabolic energy band
with effective mass $m^\ast$ and  drift velocity ${\bf v}_d$ determined by
the electron mobility and the external electric field, the electrons in the medium
are redistributed. This is determined by  a momentum shift in the electron
wave vector ${\bf k}_\parallel \to {\bf k}_\parallel -m^\ast {\bf v}_d/\hbar$
in the thermal-equilibrium  Fermi
function $f_0(\epsilon_{\bf k} )$. By making a change of variables
in the well-known Lindhard polarization function $\Pi^{(0)}_{2D}(q_\parallel,\omega)$, this effect is
equivalent to a frequency shift $\omega\to \omega-{\bf q}_\parallel\cdot {\bf v}_d$.
For massless Dirac fermions in graphene with linear energy dispersion, this Doppler
shift in frequency is not in general valid for arbitrary wave vector.
This is our conclusion after we relate
the surface    current density to the center-of-mass wave vector in a steady state.
Our calculation shows that the redistribution of electrons  leads to a shift in the
wave vector appearing in the Fermi  function by the center-of-mass wave vector
${\bf K}_0=(k_F/v_F){\bf v}_d$ where $k_F$ and $v_F$ are the Fermi wave vector and
velocity, respectively.  However, in the long wavelength limit, $q_\parallel\to 0$, the Doppler
shift in frequency is approximately obeyed.  Consequently, regardless of the nature
of the 2D layer represented in the dispersion equation we may replace
$\omega\to \omega-{\bf q}_\parallel\cdot {\bf v}_d$ in the dispersion equation in the presence
of an applied  electric field at long wavelengths.

\begin{figure}
\centering
\includegraphics[width=0.5\textwidth]{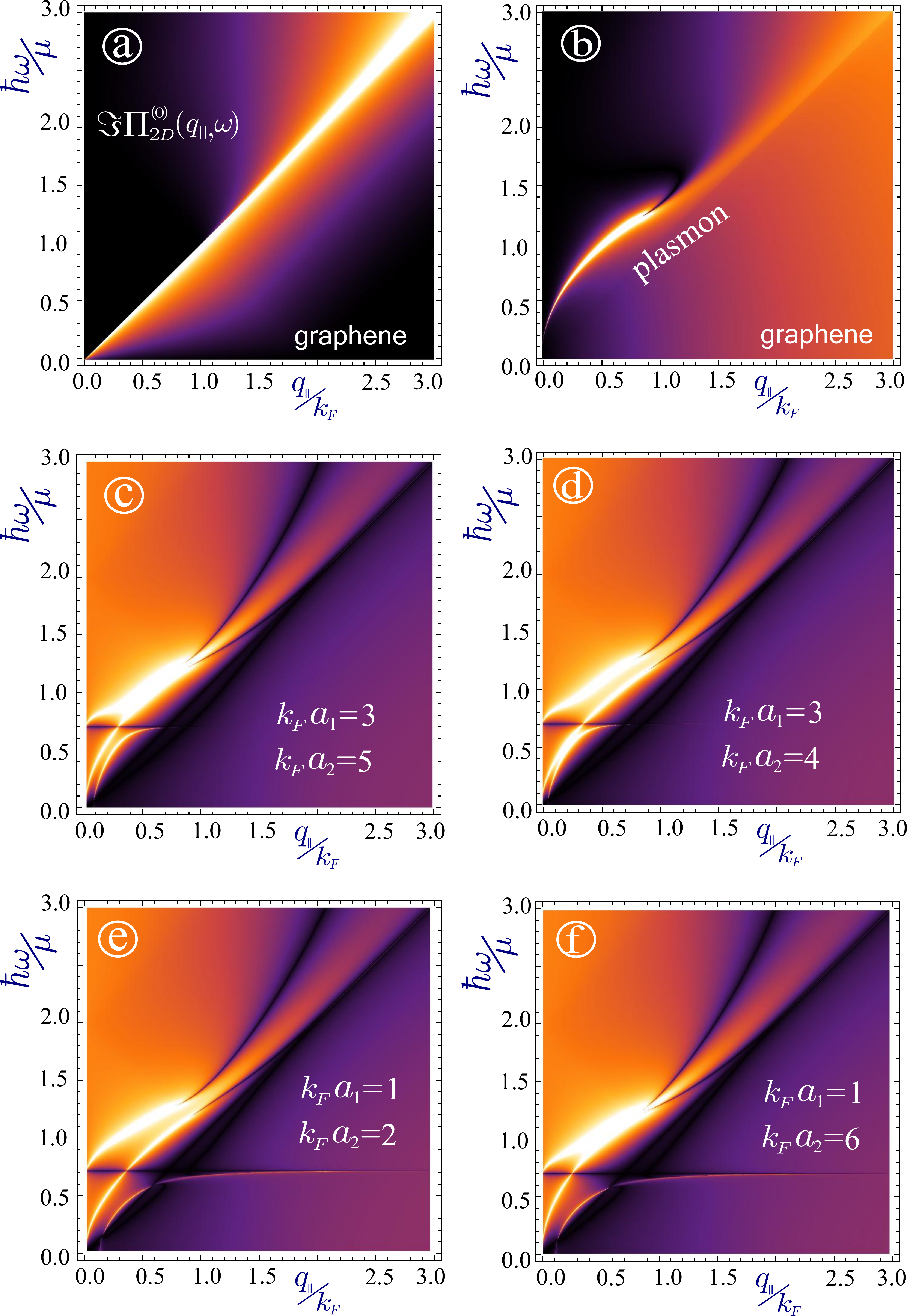}
\caption{(Color online)  Particle-hole modes and plasmon dispersion for pristine
graphene and for a hybrid system consisting of a heavily doped semiconductor interacting
with two layers. Panel $(a)$ presents the single-particle excitation spectrum, panel $(b)$
shows the plasmon spectrum  for a single layer of free standing graphene with no energy gap.
Panels $(c)$-$(f)$ show the plasmon excitations for various  chosen
distances between the surface and the layers, so that $  k_Fa_1=3 $ and $ k_Fa_2=5 $ are shown in panel $(c)$,
$k_Fa_1=3$ and $k_Fa_2 =4$ in panel $(d)$, $k_F a_1 =1$ and $k_Fa_2=6$
in panel $(e)$. Also, $k_Fa_1=1$ and $k_Fa_2=6$ are chosen in panel $(f)$.}
\label{FIG:a2}
\end{figure}

\section{Numerical Results and Discussion}

We shall treat the solution frequencies $\omega_{\pm}(q_\parallel)$ as complex variables with
${\rm Im}[\omega_{\pm}(q_\parallel)]\geq 0$, where ${\rm Im}[\omega_{\pm}(q_\parallel)]>0$ implies
finite growth rates $\gamma_{\pm}(q_\parallel)={\rm Im}[\omega_{\pm}(q_\parallel)]$ for two split plasmon modes.
Since $\epsilon(q_\parallel,\,\omega)$ is a complex function, we ask for
${\rm Re}[\epsilon(q_\parallel,\,\omega)]={\rm Im}[\epsilon(q_\parallel,\,\omega)]=0$.
Therefore, we are left with damping-free plasmon modes in the spatial domain in the system but
they still face possible instability in the time domain due to ${\rm Im}[\omega_{\pm}(q_\parallel)]>0$.

\begin{figure}
\centering
\includegraphics[width=0.5\textwidth]{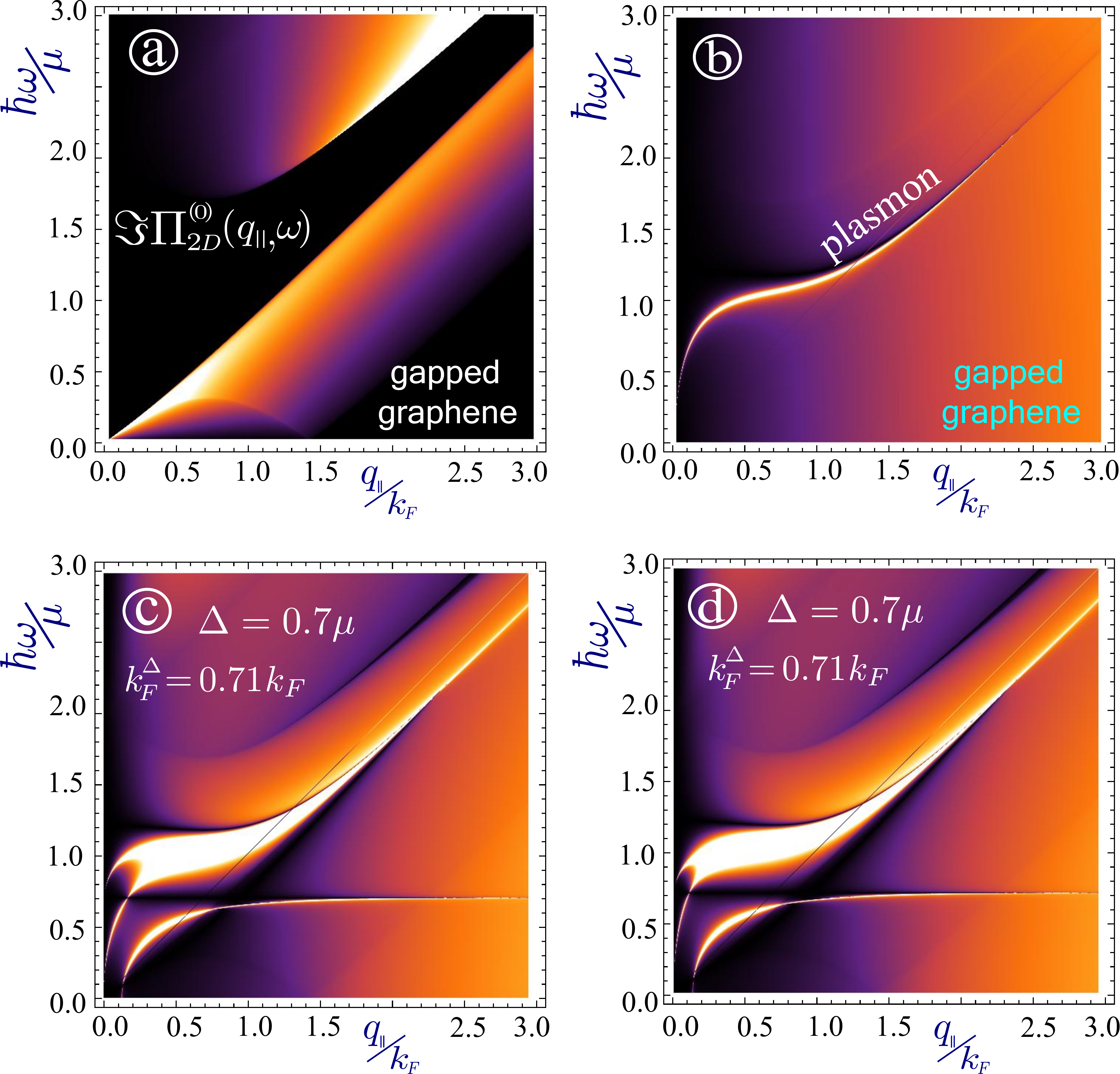}
\caption{(Color online)  Particle-hole mode region and plasmon dispersion relation
for free standing gapped graphene and for a hybrid system consisting
of a heavily doped semiconducting substrate and two layers of gapped graphene with
a large energy gap. In panel $(a)$, we show the particle-hole continuum for gapped
graphene. Panel $(b)$ shows the plasmon dispersion for a single layer of gapped graphene.
The energy gap is $\Delta = 0.7 \mu$ for all the panels, so that the Fermi momentum is
$k_F^{\Delta} = \sqrt{\mu^2 - \Delta^2}/(\hbar v_F) = 0.71 k_F^{\Delta=0}$.
Panel $(c)$ shows the case with $k_Fa_1=3 $ and $k_Fa_2=5$
and in panel $(d)$, we choose $k_Fa_1= 3$ and $ k_Fa_2=10$.}
\label{FIG:a3}
\end{figure}

\begin{figure}
\centering
\includegraphics[width=0.5\textwidth]{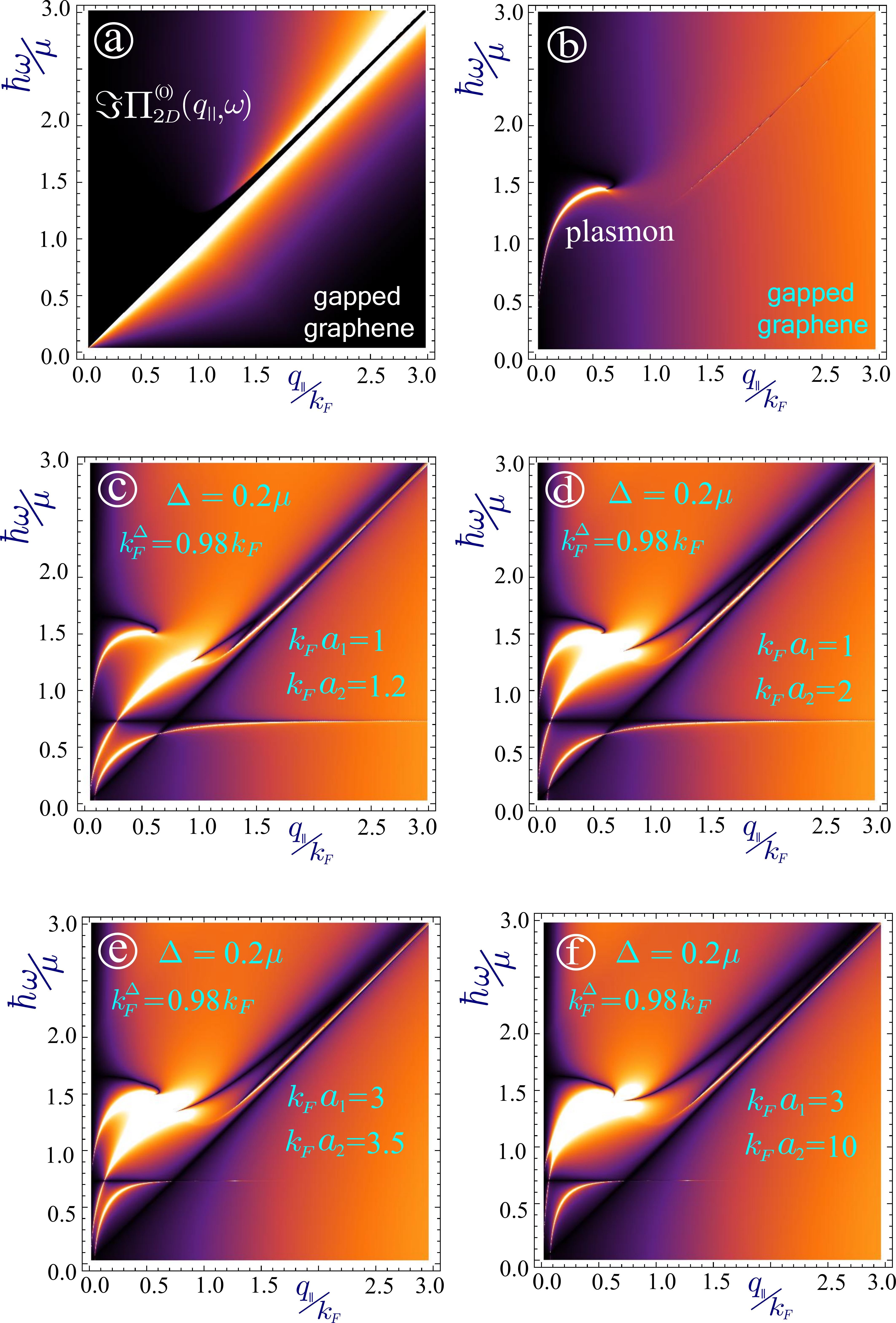}
\caption{(Color online)  Density plots of the plasmon dispersion relations for a composite system of
a semi-infinite heavily doped  semiconductor and two graphene layers, with small energy gap.
The energy gap is $\Delta=0.2\,\mu$ for all the panels, so that the Fermi momentum is
$k_F^{\Delta} = \sqrt{\mu^2 - \Delta^2}/(\hbar v_F) = 0.98 k_F^{\Delta=0}$. Panel $(a)$
shows the single-particle excitations regions, panel $(b)$ demonstrates the behavior of the
plasmons for a single layer of free standing gapped graphene. All the other panels show
the plasmon dispersion relation for the hybrid system. Panels $(c)$ and $(d)$ correspond
the choice of $k_F a_1= 1$ and $k_Fa_2=1.2$ in $(c)$ and $k_F a_1= 1$ and $k_Fa_2=2$ in $(d)$.
Panel $(e)$ shows the case of $k_F a_1=3$ and $k_Fa_2 = 3.5$, while in panel $(f)$, we choose $k_Fa_1=3$
and $k_Fa_2=10$.}
\label{FIG:a4}
\end{figure}

The number of   plasmon excitation branches for chosen $q_\parallel$
 as well as the instability domain depends on the magnitude of the drift
current and also on the distance between the constituents of the hybrid system.
In Fig: \ref{FIG:a1}, we show the complex plasma frequency solutions of our dispersion
equation, with the real parts representing the dispersion, and the imaginary part
corresponding to the rate of growth of the instability (inverse of $\Im\omega$).
For a hybrid system made up of a semi-infinite heavily doped semiconductor
and a 2D layer, we obtain two plasmon frequency solutions in the absence of a current. As
the current is increased, another solution with positive real part of the frequency may be
determined, but there may be no imaginary part of this solution until the current
reaches a critical value. The instability is closely connected to the observed
closed loop of the real solution. Mathematically, it could be explained by the fact
that for an even power equation with real coefficients for zero current all the solutions must be
complex conjugated, e.g., have  identical real parts.
\medskip

We now turn to discuss the plasmon dispersion for a hybrid   semi-infinite  conductor and
\textit{two}  layers of graphene, both with and without an energy gap. First, we investigate
the single-particle excitation region responsible for Landau damping, outside of which undamped
plasmons could be observed. The regions of the undamped plasmons is given by solving $\Im \Pi^{0}_{2D}(q_\parallel,\omega) = 0$. The results are presented in Fig.\,\ref{FIG:a2}(a)
for gapless grpahene, and in Figs.\,\ref{FIG:a3}(a) and \ref{FIG:a4}(a)
for   large ($0.7\mu$) and small ($0.2\mu$) energy band gap.
Next, we show the plasmon in a single graphene layer for each case\,\cite{DasSarma,wunsch, pavlo}.
If there is no energy gap [see Fig.\,\ref{FIG:a2}(b)] or its value is small (compared
with the chosen gap) in Fig.\,\ref{FIG:a4}(b), the plasmon is undamped only for a
small range of $q_\parallel$. In contrast, for  large energy gap, Fig.\,\ref{FIG:a3}(b)
shows that  an undapmed plasma is found for a larger range of the wave vector.
\medskip

Finally, we consider the plasmon dispersion when there are two 2D layers and a semi-infinite
conductor with free carriers. Our solutions are  determined by the location of the particle-hole
mode regions for each case, so that we obtain  longer undamped plasmon branches for the
case of gapped graphene, shown  in Figs.\,[\ref{FIG:a3}](c)  and  (d). The distance between
the surface and each of the 2D layers is also an important factor. When the larger distance
$a_2$ is chosen, the middle plasmon branch, i.e., the one which starts from the origin
with a bigger slope, tends to get close   to coincide with the the uppermost branch,
which starts from $\omega^{(0)} = \omega _p/\sqrt{2}$. For   finite energy gap, the
`surface' plasmon mode shows much brighter and broader peak, compared to the lowest
branch as depicted in  Figs.\,\ref{FIG:a3}(b)-\ref{FIG:a3}(d) and Figs.\,\ref{FIG:a4}(c)-\ref{FIG:a4}(f).\

\section{Concluding Remarks}
\label{sec4}

In summary, we are proposing a hybrid  quantum plasmonic device  which employs
2D layers in combination with a thick conducting material. We find that the spiler emits
electromagnetic radiation when a current is  passed through  the 2D layer
to make the plasmons become unstable at a specific frequency and wave number and grow over
a period of time which is determined by the positive imaginary part of its complex frequency.
It is possible to change the range of plasmon
instability by selecting the properties of the nanosheet or frequency of the
surface plasmon, i.e., the substrate.  The surface plasmon plays a crucial role
in giving rise to the Rabi-like splitting and the concomitant streams of quasiparticles
whose phase velocities are in opposite directions when the instability takes place.
The emitted electromagnetic radiation may be collected from regions on the surface that are
convenient.
\medskip

Promising sources of terahertz (THz) radiation have been investigated
over the years. These frequencies cover the electromagnetic spectrum
from microwave to infrared. Epitaxial growth of hybrid layers of   semiconductors
resulted in quantum well  structures emitting high power THz across a wide frequency range.
So far, only ultra-long wavelength emission  has been reported. Our work shows how we
 may modify this limitation with the
use of the proposed hybrid structure through  a spontaneous generation of plasmon excitations
and subsequent Cherenkov radiation  at sufficiently high draft velocities.

\acknowledgments
This research was supported by  contract \# FA 9453-13-1-0291 of
AFRL.  DH would like
to thank the Air Force Office of Scientific Research (AFOSR) for its support.


\begin{references}
\bibitem{new1}U. Banin, Y. Ben-Shahar, and K. Vinokurov,
Chem. Mater. {\bf 26}, 97 (2014).

\bibitem{new2}K. A. Brown, Q. Song, D. W. Mulder, and P. W. King, 
ACS Nano (2014) [DOI: 10.1021/nn504561v].

\bibitem{new3}P. S. Dilsaver, M. D. Reichert, B. L. Hallmark, M. J. Thompson, and J. Vela,
J. Phys. Chem. C {\bf 118},  21226 (2014).

\bibitem{new4}N. Liakakos, C. Gatel, T. Blon, T. Altantzis, S. Lentijo-Mozo, C. Garcia-Marcelot, L.-M. Lacroix, M. Respaud, S. Bals, G. Van Tendeloo, and K. Soulantica, 
Nano Lett. {\bf 14}, 2747 (2014).

\bibitem{new5}L. J. Hill, N. E. Richey, Y. Sung, P. T. Dirlam, J. J. Griebel, E. Lavoie-Higgins, I.-B. Shim, N. Pinna, M.-G. Willinger, W. Vogel, J. J. Benkoski, K. Char, and J. Pyun,
ACS Nano {\bf 8}, 3272 (2014).

\bibitem{new6}X. Ye, D. R. Hickey, J. Fei, B. T. Diroll, T. Paik, J. Chen, and C. B. Murray, 
J. Am. Chem. Soc. {\bf 136}, 5106 (2014).

\bibitem{new7}S. Blumstengel,  S. Sadofev, C. Xu, J. Puls, R. L. Johnson, H. Glowatzki, N. Koch, and F. Henneberger, 
\prb {\bf 77}, 085323 (2008).

\bibitem{new8}M. Fu, K. Wang, H. Long, G. Yang, P. Lu, F. Hetsch, A. S. Susha, and A. L. Rogach,
Appl. Phys. Lett. {\bf 100}, 063117 (2012).

\bibitem{new9}J. A Kurzman, M.-S. Miao, and R. Seshadri,
J. Phys.: Condens. Matt. {\bf 23}, 465501 (2011).

\bibitem{new10}G. Wang, X. Jiang, M. Zhao, Y. Ma, H. Fan, Q. Yang, L. Tong, and M. Xiao,
Opt. Expr. {\bf 20}, 29472 (2012).

\bibitem{Arxiv}G. Gumbs, A. Iurov, and D. H. Hunag, 
arXiv:1410.2851 [cond-mat.mtrl-sci] (2014).

\bibitem{Tso}N. J. M. Horing, H. C. Tso, and G. Gumbs, 
\prb {\bf 36}, 1588 (1987).

\bibitem{DasSarma}S. Das Sarma and A. Madhukar,
\prb {\bf 23}, 805 (1981).

\bibitem{wunsch}B. Wunsch, T. Stauber, F. Sols, and F. Guinea,
New J. Phys. {\bf 8}, 318 (2006).

\bibitem{pavlo}P. K. Pyatkovskiy, 
J. Phys.: Condens. Matt. {\bf 21}, 025506 (2009).
\end{references}
\end{document}